\documentclass[twocolumn,prl]{revtex4}

\usepackage{amsmath,amssymb}
\usepackage{epsfig}
\newcommand{\comment}[1]{}
\newcommand{\bra}[1]{\langle #1|}
\newcommand{\ket}[1]{|#1\rangle}

\newcommand{\ketbra}[2]{|#1\rangle\!\langle#2|}

\newcommand{\cpb}{CPB}
\newcommand{\upb}{UPB}
\newcommand{\eb}{EB}
\newcommand{\class}{Class}
\newcommand{\rhobb}[1]{\rho_{#1_{\mathrm{(BB84)}}}}

\newcommand{\tr}{\mathrm{tr\,}}

\usepackage{theorem}
\newtheorem{definition}{Definition}
\newtheorem{proposition}[definition]{Proposition}

\def\squareforqed{\hbox{\rlap{$\sqcap$}$\sqcup$}}
\def\qed{\ifmmode\squareforqed\else{\unskip\nobreak\hfil
\penalty50\hskip1em\null\nobreak\hfil\squareforqed
\parfillskip=0pt\finalhyphendemerits=0\endgraf}\fi}
\def\endenv{\ifmmode\;\else{\unskip\nobreak\hfil
\penalty50\hskip1em\null\nobreak\hfil\;
\parfillskip=0pt\finalhyphendemerits=0\endgraf}\fi}
\newenvironment{proof}{\noindent \textbf{{Proof~} }}{\qed}
\begin{document}
\title{``Quantumness'' versus ``Classicality'' of Quantum States}
\author{Berry Groisman$^1$, Dan Kenigsberg$^2$ and Tal Mor$^2$\\
\small 1. Centre for Quantum Computation, DAMTP, Centre for
Mathematical Sciences,\\ \small University of Cambridge, Wilberforce
Road,
Cambridge CB3 0WA, \textsc{United Kingdom} \\
\small 2. Computer Science Department, Technion,
  Haifa 32000 \textsc{Israel}\\
}

\date{\today}
\begin{abstract}

Entanglement is one of the pillars of quantum mechanics and quantum
information processing, and as a result the \emph{quantumness} of nonentangled
states has typically been overlooked and unrecognized.
We give a robust definition for the classicality versus
quantumness of a single
multipartite quantum state, a set of states,
and a protocol using quantum states.
We show a variety of nonentangled (separable) states that exhibit interesting
quantum properties, and we explore the ``zoo'' of separable states;
several interesting subclasses are defined based on their diagonalizing bases,
and their non-classical behavior is investigated.
\end{abstract}
\maketitle

\paragraph{Introduction:}

Consider an isolated discrete classical system with $N$ distinguishable
states.
The most general state of the classical system
is a probabilistic distribution over these distinguishable states.
Now consider its counterpart, an isolated discrete \emph{quantum}
system. Its most general state is a probabilistic mixture of pure
states drawn from an $N$-dimensional Hilbert space.
Yet, in various special cases, the quantum state seems to be 
identical to a classical probability distribution.
Similarly, in various special cases, a quantum protocol
using a set of quantum states seems to be practically identical to a classical
protocol which is using a classical set of states.
Our first goal is to define such special quantum states 
that are equivalent to classical probability distributions;
we also define sets of classical states and
classical protocols. 

Quantumness of states (for instance, their ``quantum correlations'')
is often associated with their entanglement, and it is
sometimes even assumed (explicitly or implicitly) that non-entangled states
can be considered ``classical''.  We argue that this is not the case,
because some (actually, most) non-entangled
states do exhibit non-classical features. Intuitively speaking,
only quantum states that correspond {\em exactly} to a classical probability
distribution can potentially be considered classical; most
nonentangled states can only be written as a probability distribution over
tensor-product quantum states, e.g., for bipartite systems
$\rho_{\rm sep} = \sum_i p_i \ket{\phi_i}_A\ket{\psi_i}_B
 \bra{\phi_i}_A\bra{\psi_i}_B$,
hence do not usually resemble any conventional distribution over classical
states. While entanglement is extensively analyzed and quantified
(see~\cite{VP:ent_m_pp,GPW05}, and references therein), the 
``quantumness'' of nonentangled (separable) states 
has typically been overlooked and unrecognized.
Our second goal is to present the quantumness exhibited by
various separable states, and to  
explore the ``zoo of separable states''.
Our last goal is to define (and make use of) measures of quantumness
${\cal Q}(\rho)$ that
vanish on any classical state $\rho_{\rm classical}$.

\paragraph{Classicality of Quantum States and Quantum Protocols:}
\label{paragraph2}

If a quantum state or a quantum protocol has
an exact classical equivalent it cannot present any interesting nonclassical
properties
nor any advantage over its analogous classical counterpart.
The state(s) of the quantum system can then
potentially be considered ``classical''.
For instance, if a single quantum system is prepared
in one of the states $\ket{0}$,  $\ket{1}$, $\ket{2}$, etc., 
and is then measured in this computational basis,
there is nothing genuinely quantum in that process. 
Tensor product states of multipartite system can also be considered classical.
Consider
a set of states in the computational basis, e.g., $\{\ket{00};
\ket{01}; \ket{10}; \ket{11}\}$; this set
has a strict classical analogue --- the classical states
$\{00; 01; 10; 11\}$. As long as no other quantum states are
added to the set (or appear in a protocol which is using these states),
the analogy is
kept, so these quantum states can be considered classical.
Tensor product states such as
$\ket{-}\ket0\ket{+}$ (where $\ket{\pm}=[\ket{0} \pm \ket{1}]/\sqrt{2}$) 
can also
be considered classical as we soon explain.

%
First, we define classical bases. We justify our claim 
that any such basis presents no quantumness, and we justify (via 
many examples) why bases that do
not follow our ``classicality'' definition are ``quantum''.

We start
with a single system and then move to bipartite and multipartite
systems:
\begin{definition}\label{class-basis-uni-part}
Let $A$ be a quantum system. Any orthonormal basis
$\{\ket i_A\}$ of $A$ 
can be considered as a \emph{classical basis} of
the system.
\end{definition}
For example, the computational basis
$\{\ket{0};\ket{1}\}$
of a single qubit is obviously classical.
The Hadamard basis $\{\ket{+};\ket{-}\}$ is also classical.

One may argue that our definition is too flexible and that Nature
allows only one basis to be classical.
For instance an alternative for Def.~\ref{class-basis-uni-part} is
\begin{description}
\item
Let $A$ be a quantum system with a single \emph{preferred} orthonormal basis
$\{\ket i_A\}$, in the sense that measurements
can only be performed in this basis.
Only this basis
can be considered as a \emph{classical basis} of
the system.
\end{description}
We do not agree to that narrower definition.
First, nothing in conventional quantum
theory favors one of the system's bases over any other.
Second, although in the more general
\emph{relativistic quantum field theory} it is commonly
believed that Nature generally provides a preferred
basis, 
on time-scales sufficiently short (e.g., short enough for performing
quantum computation), all bases are equivalent.

We now move to defining classical bases for bipartite and
multipartite systems.
\begin{definition}\label{class-basis-bi-part}
Let $A$ and $B$ be two quantum systems with orthonormal bases
$\{\ket i_A\}$ and $\{\ket j_B\}$ respectively. The 
tensor-product basis
$\{\ket i_A\otimes\ket j_B\}_{ij}$ is a \emph{classical basis} of
the bipartite system.
\end{definition}
\begin{definition}\label{class-basis-multi-part}
(recursive)
Let $A$ be a (bipartite or 
multipartite) quantum system with a \emph{classical} basis $\{\ket
i_A\}$, and let $B$ be a unipartite quantum system  with an
orthonormal basis $\{\ket j_B\}$.
The tensor-product basis $\{\ket i_A\otimes\ket j_B\}_{ij}$
is a \emph{classical basis} of
the composite $AB$ system.
\end{definition}
The redundancy in Defs.~\ref{class-basis-bi-part}--\ref{class-basis-multi-part}
is kept for readability.

Let us see a few examples.
For two qubits, the computational basis is classical, as well as the basis
$\{\ket{++};\ket{+-};\ket{-+};\ket{--}\}$.
On the other hand, the Bell basis $\{\ket{\Phi_\pm};\ket{\Psi_\pm}\}$ is obviously non-classical, and
more interestingly, even
the basis
$\{\ket{00};\ket{01};\ket{1+};\ket{1-}\}$
is non-classical, too.

Having identified classical bases, we proceed to define a
classical state and a set of classical states.
\begin{definition}\label{defined-class-state}
A state $\rho$ is a classical state, iff there exists a classical
basis $\{\ket{v_i}\}$ in which $\rho$ is diagonal.
\label{class_state}
\end{definition}
Following our definition,
any (single) state $\rho$ (either pure or mixed) of a single system $S$ 
\emph{can always be considered classical}.
A joint state of two or more quantum systems
can also either be pure or mixed.
If it is pure it is either a tensor product state
or an entangled state.
Following the classicality definitions, any such tensor-product state
is classical while any such entangled state is nonclassical.
For mixed bipartite or multipartite states the situation is much more
complicated:
Tensor-product mixed states are obviously still classical
as each subsystem can be diagonalized in a classical basis of its own.
Entangled mixed states are obviously nonclassical.
Between these two extremes we can find a zoo of
separable---yet quantum---states.

We made this definition independently of a similar definition due  
to Ref.~\cite[see Sec.~5]{HHHOSSS}; they use 
the name ``(properly) classically correlated states''
which is more precise, yet longer, than our term
``classical states''.

Prior to dealing with separable quantum states we provide
two additional useful definitions.
\begin{definition}
A set of states $\rho_1 \ldots \rho_k$ is a classical set
iff all $\rho_i$ are diagonalizable in a single classical
basis.
\label{class_set}
\end{definition}
If a quantum protocol (be it computational, cryptographic, or any other physical
process) is limited to a classical set of states,
the process has an exact classical equivalent, and cannot present any advantage
over an analogous classical protocol.
More formally:
\begin{definition}
\label{classprot} A protocol (in quantum information processing)
is 
\emph{classical} iff all states involved in it belong to a single classical
set of states.
\end{definition}
If a protocol involves two or more pure nonorthogonal states it cannot be
considered classical [see~\cite{FuchsSasaki03} for a thorough analysis of
the quantumness of protocols involving only pure states.]
Yet following our definitions, even protocols involving only pure orthogonal
product-states might be highly quantum;
and similarly, even a single bipartite mixed separable 
state can be highly nonclassical.

\paragraph{Nonclassicality of Separable States:}
\label{paragraph3}
Let us prove the quantumness of several interesting separable states.

1.---
Pseudo-pure states.

A state of the form $\epsilon\ketbra\psi\psi+\frac{1-\epsilon}NI$
is called a pseudo-pure state (PPS) as the part with the coefficient
$\epsilon$ transforms as if the state was a pure state.
PPSs
focus wide interest based on theoretical and experimental grounds.
It has been shown~\cite{Popescu99} that
there is a volume of \emph{separable} PPSs
around the totally-mixed
state $I/N$; every PPS with low-enough $\epsilon$ is separable.
This fact was even used to argue
that experiments which
produce such low-$\epsilon$ states are not truly quantum.
It was later argued, however, that
albeit being separable, these states do exhibit non-classical
effects~\cite{BBKM02}. Using our definitions we see that:
\begin{proposition}
\label{pps_qness}
A PPS $\rho_\epsilon=\epsilon\rho+\frac{1-\epsilon}NI$ is quantum iff $\rho$ is,
for any $\epsilon>0$.
\end{proposition}
\begin{proof}
Any diagonalizing basis of $\rho_\epsilon$ also diagonalizes $\rho$,
independently of $\epsilon$. Since $\rho$ is quantum, it is not diagonalizable
in a classical basis, and so is $\rho_\epsilon$.
\end{proof}
This is true for any system dimension. As a special case for $N=4$,
a separable Werner state~\cite{Werner89}
\(
\chi =
\epsilon \ketbra{\Psi_-}{\Psi_-}+ \frac{1-\epsilon}{4}I
\)
is nonclassical for any $0<\epsilon\leq \frac13$ 
(see also~\cite{OllivierZurek01} for a different demonstration
of nonclassicality of the Werner states).
Note that the Werner state is also separable
and nonclassical for 
any $-\frac13 \leq \epsilon <0$.

2.---
States used for quantum key distribution.

The original quantum key distribution protocol, the BB84 protocol, involves
qubits of four different states: $\ket0$, $\ket{1}$, $\ket+$, and $\ket-$,
sent from Alice to Bob.
The protocol may also be described in a less conventional
manner~\cite{Mor98}, where Alice sends in two steps
either the state
$\rhobb0
=\frac12\left[\ketbra{00}{00}+\ketbra{1+}{1+}\right]$
to represent `0' or
$\rhobb1
=\frac12\left[\ketbra{01}{01}+\ketbra{1-}{1-}\right]$
to represent `1'; the right-hand-qubit is sent first and the
left-hand-qubit is sent later on in order to reveal the basis of the first
qubit.
\begin{proposition}
$\rhobb0$
is not classical; so is
$\rhobb1$.
\end{proposition}
\begin{proof}
Any diagonalizing product basis of $\rhobb0$ includes
$\ket0_A\otimes\ket0_B$ and $\ket1_A\otimes\ket+_B$. That basis cannot be
classical, as Bob's parts, $\ket0_B$ and $\ket+_B$, are not orthogonal and hence
cannot be members of a single classical basis. The same
reasoning applies to $\rhobb1$, too.
\end{proof}
Thus, although all the four states involved in the protocol
$\ket{00}$, $\ket{1+}$, etc.
are mutually orthogonal tensor-product
states, the protocol is highly ``quantum''.

3.---
States that present nonlocality without entanglement.

Various sets of states proposed
in~\cite{nonloc.noent99,UPB99} define
processes that exhibit nonlocal quantum behavior although none of
the participating states is entangled. In particular, spatially
separated parties cannot reliably distinguish between different
members of the set (albeit comprising of mutually orthogonal direct product
states!) without assistance of entanglement. 
For instance, the set  
$\{\ket{01+};\ket{1+0};\ket{+01};\ket{---}\}$ is nonclassical.

4.---
The Bernstein-Vazirani Algorithm.

The Bernstein-Vazirani algorithm~\cite{BV97}
generates no entanglement (see~\cite{Meyer00soph}).
However, it is clearly a quantum algorithm, with no classical equivalent.
It makes use of states from the computational \emph{and} Hadamard bases,
which are not simultaneously diagonalizable in a single classical basis.

\paragraph{A Zoo of Separable States:}
\label{paragraph5}
Within the set of all  
separable states we identify some interesting subsets 
based on their diagonalizing bases.

First let us consider the classical states $\class$, 
the states diagonalized in a classical basis:
A bipartite state 
(this argument easily extends to multipartite states)
is classical if, and only if,
Alice and Bob can perform a measurement in its (classical) diagonalizing
basis via local orthogonal measurements, without exchanging any message
(classical or quantum) and without disturbing the state.
%

The notion of diagonalizing basis is now used to define more subsets
of the separable states.
Ref.~\cite{UPB99} defines a \emph{complete product basis} (\cpb) as follows:
A \cpb{} is a complete orthonormal basis of a
multipartite Hilbert space,
where each basis element is a (tensor) product state.
We define the set of \cpb-states as follows:
\begin{definition}
A state $\rho$ is a
\cpb{}-state iff it is diagonalizable in a \cpb.
\end{definition}
Clearly, all classical states 
are \cpb{} states; but not vice
versa.
Thus, in a multipartite finite-dimensional Hilbert space
$\class \subset 
\cpb\subset SEP\subset {\cal H}_{\rm total}$.
For example, 
$\rhobb0$ and
$\rhobb1$ 
are nonclassical \cpb{}-states diagonalized in the \cpb{} 
$\{\rho_{00};\rho_{01};\rho_{1+};\rho_{1-}\}$.
Note that local operations   
and unidirectional 
classical communication, 
but {\em without} adding the ability to ``forget'', 
are sufficient for converting the BB84 states into
classical states.
These operations are a very special case of 
the well-known LOCC (local operations and classical
communication) that {\em include the ability to forget}, and that are 
therefore sufficient for generating any separable state.
A slightly more complicated (qubit plus qutrit) state, $\rho
=\frac13\left[\ketbra{00}{00}+\ketbra{1+}{1+}+\ketbra{+2}{+2}\right]$
requires 
local operations (again, without ``forgetting'') and bidirectional 
classical communication in order for it to be converted into a classical state.
We call these two types of \cpb-states ``unidirectional \cpb-states''
and ``multi-directional \cpb-states'' respectively.

Interestingly, there are \cpb-states that belong to neither subsets:
consider a state built 
from a probability distribution 
over {\em all} the eight states~\cite{nonloc.noent99} 
$\{\ket{01\pm};\ket{\hbox{$1\!\pm\!0$}};\ket{\hbox{$\pm01$}};\ket{000};\ket{111}\}$;
although it is a \cpb-state,
such a state cannot be converted into a classical states unless
quantum communication is allowed, 
or unless the general LOCC
(including the power of ``forgetting'') are allowed.
Thus, we specify also a third subset of the \cpb{} states ---
``Q-convertible \cpb-states''.

Let $V$ be an orthonormal basis of a 
subspace of a multipartite Hilbert space $\cal H$,
where each basis element is a (tensor) product state.
Ref.~\cite{UPB99} defines that  
$V$ is an \emph{unextendible product basis} (\upb) if
the subspace ${\cal H} - \mathrm{span}\{V\}$ contains no product state.
We define the set of \upb-states as follows:
\begin{definition}
A separable state $\rho$ is a 
\upb{}-state iff it is diagonalizable in a \upb.
\end{definition}
Note that a \upb-state~\cite{UPB99} such as
\(
\rho_{\epsilon\upb}=(1-6\epsilon)\rho_{01-}+\epsilon\rho_{1-0}+2\epsilon\rho_{-01}+3\epsilon\rho_{---}
\label{exampleupb}
\), 
proves that there are \upb-states that are not in \cpb{}.
Note also that with $\epsilon\rightarrow0$, this state  
is infinitesimally close to a classical state.
More relations and borderlines between these sets and  
also the set \eb{} (see below) will be explored in future 
research~\cite{GKM-future}.

We identified another class of \upb-states 
that can be proven to be non-classical:
\begin{proposition}
The uniform mixture of UPB elements
$\rho_{\rm UPB} = (\rho_{01+}+\rho_{1+0}+\rho_{+01}+\rho_{---})/4$
is nonclassical.
\end{proposition}
\begin{proof}
Assume that $\rho_{\rm UPB}$ is classical. The same classical basis that
diagonalizes it, also diagonalizes the state $I/4-\rho_{\rm UPB}$.
However, this
contradicts the fact that it is 
\emph{bound-entangled}~\cite{UPB99}
and therefore quantum.
\end{proof}

The last set we define is the set \eb{} of states diagonalized 
only in a non-product basis.  
As we had already seen, many separable states belong to this \eb{} set,
e.g., various PPS and Werner states. Obviously, 
all non-separable states also belong to
this set.

\paragraph{Measures of quantumness:}\label{Section4}
A measure of nonclassicality (quantumness), ${\cal Q}(\rho)$, of a
state $\rho$ has to satisfy two conditions; (a) ${\cal
Q}(\rho)=0$ iff $\rho$ is classical, 
(b) ${\cal Q}(\rho)$ is invariant under local unitary operations.
One might also expect a third condition;
(c) ${\cal Q}(\rho)$ is monotonic under local operations (without
classical communication)
\footnote{These conditions resemble the line of thought used in
searching for the measure of entanglement \cite{VP:ent_m_pp}.};
Yet, condition (c) is not always satisfied by quantum states: The 
classical state
$\frac12\ketbra{00}{00}+ \frac12\ketbra{03}{03}$ of a $2\times 4$ 
system can be converted to $\rhobb0$ just by the power of
forgetting---Bob redefines his qu-quadrit as two qubits
with $\ket0_{\rm quad} = \ket{00}$ and $\ket{3}_{\rm quad} = \ket{1+}$, and forgets 
his first qubit.

A class of measures of quantumness of $\rho$ is defined as
\begin{equation}
\label{distance-classical}
{\cal Q}_D(\rho)=\min_{\rho_c}D(\rho,\rho_c)\end{equation}
where $D$ is any measure of distance between two states such that
the conditions (a)-(b) are satisfied, and
the minimum is taken over
all classical states $\rho_c$. 
One of the natural candidates for $D$
is the relative entropy 
$S(\rho\|\rho_c)=\tr\rho\log\rho-\tr\rho\log\rho_c$, in which case
we refer to it as
${\cal Q}_{\rm rel}(\rho)$ --- the \emph{relative entropy of
quantumness}.
The benefit of using the relative entropy as a measure
is that it was extensively studied for measuring 
entanglement~\cite{VP:ent_m_pp} (relative 
to the closest separable state).
Thus, we can adopt and make use of some known results, and we can also 
monitor the connection between the quantumness of states 
and their entanglement.
Other measures (or their variants) 
that can potentially be very useful are the  
\emph{fidelity of quantumness} and \emph{Von Neumann mutual information}
that will be explored in future research~\cite{GKM-future}.

%
For bipartite pure states,
the relative entropy of quantumness
equals its entropy of entanglement. 
In other words, a pure
state is quantum as much as it is entangled.
Any bipartite entangled state $\ket{\Psi}$  can be written in a
Schmidt decomposition
$\ket{\Psi}=\sum_{i=1}^{d}c_i \ket{ii}_{AB}$,
where $c_i \geq 0$ 
and $d=\min[d_A,d_B]$, $d_A$,
$d_B$ are dimensions of local Hilbert spaces.
If we use the relative
entropy of entanglement 
then 
the closest separable state~\cite{VP:ent_m_pp} is
\begin{equation}
\sigma_{cl}=\sum_{i=1}^{d}(c_i)^2\ketbra{ii}{ii}_{AB}.
\end{equation}
This state happens to be also classical,
and thus the relative entropy of quantumness 
(which is equal to its relative entropy of entanglement) is 
${\cal Q}_{\rm rel}(\Psi)=-\sum_i(c_i)^2\log[(c_i)^2]$. 
[The classical state $\sigma_{cl}$ lies on entangled-separable boundary.]
Note that the 
quantumness of a
maximally entangled state 
is ${\cal Q}_{\rm rel}(\Psi_{ME})=\log d$.

%
Let us present some mixed states for which their quantumness can easily
be calculated:
According to~\cite[Th.~4]{VP:ent_m_pp}, 
$\sigma_{cl}$ is the separable state that minimizes
$S(\rho_p\|\sigma_{cl})$ for any state of the form
$\rho_p=p~\ketbra{\Psi}{\Psi}+(1-p)\sigma_{cl}$, too.
Therefore, the relative entropy of entanglement of $\rho_p$
equals to its relative entropy of quantumness.
%

Given any bipartite state $\rho_{AB}$, let its \emph{Schmidt basis} be the
(classical) basis diagonalizing $\tr_B\rho_{AB}\otimes\tr_A\rho_{AB}$. Let
$\rho_{\rm Sch}$ be produced from $\rho_{AB}$ by writing it 
in its Schmidt basis and having all off-diagonal
elements zeroed.  The state $\rho_{AB}$ and its Schmidt state yield
identical classical correlations if measured in the Schmidt basis.

The Schmidt state can be found very useful for 
defining quantumness for 
any state $\rho_{AB}$, as $\rho_c$ is usually unknown;        
instead of using Eq.~(\ref{distance-classical}) as a measure, one can directly
refer to the distance between a state $\rho_{AB}$ and its corresponding 
Schmidt state:
\begin{equation}
\label{distance-Schmidt}
{\cal Q}_D(\rho)=D(\rho_{AB},\rho_{\rm Sch})\end{equation}
as a measure of quantumness of a state.
If we now use the relative entropy,
the resulting measure satisfies confitions (a) and (b).

We saw above, that for a pure bipartite state the Schmidt state 
$\rho_{\rm Sch} = \sigma_{cl}$ is the closest
classical state. 
One might conjecture that for any bipartite state $\rho$, 
the closest classical state (using
relative entropy measure) is its Schmidt-state $\rho_{\rm Sch}$.
This however, is not true. For instance, 
we checked the \cpb-state $\rhobb0$ which is useful in quantum key 
distribution;  
it is interesting to note that 
either the classical state 
$\frac{1}{2}\rho_{00}+\frac{1}{4}\rho_{10}+\frac{1}{4}\rho_{11}$
or the classical state 
$\frac{1}{4}\rho_{0+}+\frac{1}{4}\rho_{0-}+\frac{1}{2}\rho_{1+}$,
are actually closer to
$\rhobb0$ than 
its Schmidt state --- a state diagonal in the classical basis
(known as the Breidbart basis)
$\{\rho_{0b_0};\rho_{0b_1};\rho_{1b_0};\rho_{1b_1}\}$ (where
$\ket{b_0}=\cos{\frac\pi8}\ket0-\sin{\frac\pi8}\ket1$,$\ket{b_1}
=\sin{\frac\pi8}\ket0+\cos{\frac\pi8}\ket1$).
It is easy to verify numerically that the above two states are  
the closest ones to $\rhobb0$, hence can be used for calculating its
relative entropy of quantumness.  
We also proved this fact analytically but the proof
is too long and therefore is not included in this Letter,
but is left for an extended paper~\cite{GKM-future}.
The entropy of quantumness relative to the Schmidt state is different in this
case of course.

\paragraph{Summary:}
In this Letter we gave definitions for classical states and protocols in 
quantum information processing. We explored the ``zoo'' of separable states,
we gave a good number of examples 
and we defined some useful measures for the quantumness of non-classical
states. 
Our measures and our analysis are mainly based on the notions of 
``diagonalizing basis'' and the ``Schmidt basis'' (which are identical in the
case of pure entangled states).
Other measures of quantumness have been defined and used previously: 
Ref.~\cite{OllivierZurek01} defines the \emph{quantum
discord} between the parts of a
bipartite state. 
Ref.~\cite{HHHOSSS} extensively uses  
the \emph{quantum information deficit} measure of quantumness,
and the relative entropy of quantumness (which we use independently).
Section~5 in~\cite{HHHOSSS} provides very interesting 
subclasses --- yet, different from ours --- of the separable states. 
Their class of ``informationally
nonlocal'' states seems to be identical to our two subclasses ---
the \upb-states and the unconvertible \cpb-states.

TM and BG are grateful to Amit Hagar and the workshop he organized on
``What is quantum in quantum computing'' (Konstanz, May 2005);
This collaborative effort (mainly, 
Defs.~\ref{class-basis-bi-part}--\ref{defined-class-state})
began there, with the workshop as its inspiration.


\end{document}